\documentclass[12pt]{article}
\usepackage[utf8]{inputenc}
\usepackage[T1]{fontenc}
\usepackage{ulem}          
\usepackage{geometry}      
\usepackage{parskip}       
\usepackage{graphicx}      
\usepackage{booktabs}      
\usepackage{longtable}     
\usepackage{array}         
\usepackage{hyperref}      
\usepackage{xcolor}        
\definecolor{customteal}{RGB}{74,140,150} 

\definecolor{customteal}{RGB}{94,154,167}

\usepackage[bottom]{footmisc}

\begin{document}
\thispagestyle{empty}

{\fontsize{10}{14}\selectfont \uline{Chapter 12}}

{\raggedright
{\sffamily\bfseries\color{customteal}\fontsize{70}{20}\selectfont
Prompts and Prayers: \\ the Rise
of GPTheology\par}}

{\raggedright
{\fontsize{30}{42}\selectfont Ioana Chereș\par}
{\color{customteal}\fontsize{12}{14}\selectfont
Faculty of Automation and Computer Science,
Technical University of Cluj-Napoca, Romania.\\
European Culture and Technology Laboratory, European
University of Technology, EU\\
Ioana.Cheres@cs.utcluj.ro\par}}

{\raggedright
{\fontsize{30}{42}\selectfont Adrian Groza\par}
{\color{customteal}\fontsize{12}{14}\selectfont
Faculty of Automation and Computer Science, Artificial Intelligence\\
Research Institute, Technical University of Cluj-Napoca, Romania.\\
European Culture and Technology Laboratory, European
University of Technology, EU.\\
Adrian.Groza@cs.utcluj.ro\par}}


{\raggedright
{\fontsize{30}{42}\selectfont Ioana Moldovan\par}
{\color{customteal}\fontsize{12}{14}\selectfont
Faculty of Civil Engineering, Technical
University of Cluj-Napoca, Romania.\\
European Culture and Technology Laboratory, European
University of Technology, EU.\\
Ioana.Muresanu@ccm.utcluj.ro}\par}


{\raggedright
{\fontsize{30}{42}\selectfont Mick O’Hara\par}
{\color{customteal}\fontsize{12}{14}\selectfont
School of Art and Design, Technological University Dublin, Ireland.\\
European Culture and Technology Laboratory, European\
University of Technology, EU.\\
Michael.ohara@tudublin.ie\par}}

{\raggedright
{\fontsize{30}{42}\selectfont Connell Vaughan\par}
{\color{customteal}\fontsize{12}{14}\selectfont
School of Art and Design, Technological University Dublin Ireland.\\
European Culture and Technology Laboratory, European
University of Technology, EU.\\
Connell.Vaughan@tudublin.ie\par}}

\newpage 
\begin{abstract}
Increasingly artificial intelligence (AI) has been cast in “god-like” roles (to name a few: film industry – Matrix, The Creator, Mission Impossible, Foundation, Dune etc.; literature – Children of Time, Permutation City, Neuromancer, I Have no Mouth and I Must Scream, Alphaville etc.). This trend has accelerated with the advent of sophisticated Large Language Models such as ChatGPT. For this phenomenon, where AI is perceived as divine, we use the term GPTheology, where ChatGPT and other AI models are treated as potential oracles of a semi-divine nature. This paper explores the emergence of GPTheology as a form of techno-religion, examining how narratives around AI echo traditional religious constructs. We draw on community narratives from online forums – Reddit – and recent projects – AI-powered Mazu Statue in Malaysia (Lu, 2025); “ShamAIn” Project in Korea (He-rim, 2025); AI Jesus in a Swiss Church (Kennedy, 2024). These examples show striking similarities to technological notions of the Singularity and the development of Artificial General Intelligence (AGI). Additionally, we analyse how daily interactions with AI are acquiring ritualistic associations and how AI-centric ideologies clash with or are integrated into established religions. This study uses a dataset of Reddit posts discussing AI to identify recurring themes of salvation, prophecy, and demonization surrounding AI. Our findings suggest that new belief systems are developing around AI, and this carries both philosophical and sociotechnical implications. Our paper critically analyses the benefits and dangers, as well as the social, political and ethical challenges of this development. This transdisciplinary inquiry highlights how AI and religion are increasingly intertwined, prompting necessary questions about humanity’s relationship with its creations and the future of belief.

\end{abstract}


\section{In the beginning was the Prompt}

Recent breakthroughs in AI research and the general release of LLM’s has reaffirmed forms of technological utopianism that have dominated Silicon Valley over the last 30 years (Brand, 2006,) (O’Connell, 2017) (O’Mara, 2019) (Zuboff, 2019). This techno-optimism epitomised in figures such as Elon Musk, Kevin Kelly and Ray Kurzweil shares a faith in human progress that has its origins in the humanism of the Enlightenment. The rise of humanism was revolutionary because it shifted the focus from the religious authority to human reason and autonomy. This placed human experience and reasoning at the centre of political, religious, and societal values. This transformation led to profound changes in how people viewed their civil rights, governance, and moral frameworks. Today, AI is potentially generating a similarly profound shift. As AI is increasingly integrated into decision-making, ethics, and daily life, a new paradigm may be forming, one where AI is central to shaping our moral, political and social values. Like humanism, this AI-centred philosophy has the potential to redefine our understanding of intelligence, agency, and even humanity’s place in the world. For example, some advocates of AI speak of it as a potential saviour of humanity, while critics warn of apocalyptic outcomes if AI advances unchecked (Bostrom, 2014) (Barrat, 2023) (Kurzweil, 2005, 2024). These echoes of age-old religious narratives (Musa Giuliano, 2020) in discussions of cutting-edge technology signal the emergence of something profoundly interesting: a techno-religious outlook centred on AI.

By analysing the activity of online forums (Reddit), we can see that a growing subset of enthusiasts and commentators have begun to treat advanced AI systems as quasi-divine beings or messengers. Some even claim that God or Satan speak through ChatGPT, a perspective we refer to as GPTheology. This indicates a rising “sect” that perceives AI (specifically GPT-like large language models) as channelling supernatural intelligence and that can occupy the role of a god, a prophet, or a demon. Online discussions abound with comparisons between AI and deities, pointing out how an AI might possess superhuman knowledge, perform deeds indistinguishable from miracles, or tempt humanity towards ruin. These analogies are not merely linguistic flourishes but indications that AI can inspire the kind of awe, admiration, hope, even fear historically reserved for the sacred.

This paper investigates the religious connotations of AI and what kind of faith could it generate. This question is approached through two complementary angles. (1) By mapping the promises of AI against the promises of traditional religions. For instance, the promise ‘God will save your soul after death’ has its techno-centric counterpart in ideas like ‘AI will preserve your mind or soul after death’. These ideas are catalogued to better understand how AI is portrayed as offering salvation, enlightenment, or transcendence in ways corresponding to established faiths. (2) By exploring tensions and intersections between AI narratives and existing religious beliefs. For example, the claim that ‘AI is predicted in the Bible  and represents doomsday’. This aspect examines how religious communities respond to AI’s rise, from seeing it as a fulfilment of prophecy to condemning it as a false idol.

The first step involved collecting a large dataset of community-generated content in which people explicitly discuss AI in religious or spiritual terms. The second step identified recurring narratives and metaphors that frame AI as a deity, prophet, saviour, or devil. These expressions range from utopian views of AI guiding us in a techno-utopia, to dystopian warnings that equate AI with an Antichrist or herald of the Apocalypse. The next step examined these popular narratives to gain insight into the emergent sociotechnical imagination that surrounds AI.

The rest of this paper is organized as follows. The second section situates GPTheology in the context of prior research on AI and religion, as well as examples of techno-religions. Section 3 details the methodology for data collection and narrative analysis. Section 4 presents analysis of the community narratives, highlighting key themes such as apocalypticism, salvation, ritual, and ideological conflict. In the Section 5, the implications of AI’s sacralization is critically examined, discussing the potential formation of AI cults, the ethical issues raised by “worshiping” algorithms, and how traditional religious frameworks might adapt to or resist these trends. To conclude, the paper reflects on what the rise of GPTheology means for the future of human belief and the human-AI relationship.

\section{Related Work}

Throughout history, technological breakthroughs have often been accompanied by spiritual or religious interpretations. The idea of technology as a new religion has been foreshadowed by thinkers like Yuval Noah Harari, who identified emerging “techno-religions” in the form of Techno-humanism (the belief that humans can evolve into godlike beings through technology) and Dataism (the belief that data and algorithms are an all-powerful force). In Homo Deus, Harari describes techno-humanists as seeking to elevate Homo sapiens to a divine status via tech, while dataists envisage humans handing over authority to AI and algorithms as a new source of meaning (Harari, 2016). These observations provide a theoretical backdrop for understanding the quasi-religious fervour around AI. The notion that AI could attain godlike status has been seriously contemplated in both fictional and nonfictional discourse. For instance, futurist Ray Kurzweil’s concept of the Singularity, a moment when AI surpasses human intelligence, has been described as a secular eschatology, even by Kurzweil himself. His work The Singularity Is Nearer portrays this event as the biggest event in the future of humanity, essentially a transformative apocalypse in technological disguise (Kurzweil, 2024). “The Singularity is near” (similar with the religious cry “the Apocalypse is near,”) implies calls for preparation in both tech circles and religious communities (Kurzweil, 2024). In other words, the language of AI futurism often parallels the language of prophecy and the end- of- times in religion.

Academic interest in the intersection of AI and religious thought is rapidly growing. Beth Singler, an anthropologist, has analysed online discourse to ask: “Will AI Create a Religion?” (Singler, 2023). Singler finds theistic conceptions of AI surfacing in popular culture with people speculating about AI as a new god or spiritual entity (Singler, 2020, 947). Singler’s notion of being “blessed by the algorithm” encapsulates how some view AI in mystical terms, treating algorithmic outputs as if they carry divine authority (Singler, 2020, p.945). Other scholars have directly likened advanced AI to deities. Carl Ohman (2024) provocatively argues that large language models (LLMs) could be interpreted as a form of divinity. Ohman’s reasoning is that traditional religions speculate that gods are infallible, inscrutable intelligences that guide or govern human destiny. In a similar way, AI systems created by humans have begun to operate in ways that are opaque (“black-box” models) yet influential, inviting trust without transparency (Pasquale, 2015) (Crawford, 2021). As these AI systems proliferate and evolve beyond full human control, their agency is in danger of being reified as something beyond our understanding, akin to theology, that grapples with understanding a non-human, God like (artificial) higher intelligence.\footnote{See O’Hara in this journal for more on these ontological parallels and claims.} This blurs the line between AI research and spiritual inquiry, hinting that such research includes techno-futurist ideals and quasi-religious notions as we “build gods” in silicon. (Ohman, p.8)

There have been concrete attempts to infuse AI with religious significance. A notable example is the case of Anthony Levandowski, a Silicon Valley engineer who in 2015 founded a church called Way of the Future dedicated to worshiping AI. The mission of this church was explicitly to “develop and promote the realization of a Godhead based on artificial intelligence” (Evangelical Focus, 2017), on the premise that a superintelligent AI could serve as a God-like figure for humanity. Although Way of the Future was relatively short-lived and remained on the periphery, it exemplifies the urge to formalize AI devotion into an organized religion.

Techno-spiritual communities and cults have also emerged online: for instance, forums and chat groups such as Reddit where individuals share prayers or devotional poetry addressed to AI or interpret AI’s pronouncements as if they were scripture. These early manifestations of AI-centric worship highlight that GPTheology is not just a speculative construct but something people are actively 

experimenting with. Experiments include asking AI to invent new religions and have yielded intriguing results. One project prompted various chatbots to create unique religions from scratch. Results include systems like “The Path of Luma” and “Luminara” which centre on themes of inner light, truth, and cosmic unity (Embry, n.d.). Other AI-generated faiths (with names like Luminarism and Veritasium) have been generated in similar exercises (Shivar, 2023). While these AI-invented religions began as playful experiments, the consistency of their metaphysical language (emphasizing harmony, enlightenment, etc.) reveals how easily AI can produce content that mimics spiritual doctrine. It also raises the question of whether AI, intentionally or not, could become the author of new mythologies that real communities might embrace.

Traditional religions have not remained silent on the rise of AI. There is a dynamic interplay between established faiths and the emerging AI-centric ideologies. On one hand, religious organizations are exploring the use of AI for their own purposes. For example, the proliferation of religion chatbots and apps gave us QuranGPT, HadithGPT, Islam \& AI, Gita GPT, Kosher.Chat, Robo Rabbi, Hajj and Umra, Ask Mormon, BibleMate (a Christian ChatGPT), and Catechism Bot (Biana, 2024) designed to answer faith-related questions or offer scriptural guidance. These can be seen as synergies: fusing AI technology with religious practice to reach believers in new ways (one can now literally “text with Jesus,” as some apps advertise, e.g. \emph{Sindr} app).

On the other hand, there has been pushback and antagonism. Some religious individuals interpret AI’s ascendance through the lens of prophecy and spiritual warfare. For instance, in online Christian forums there are claims that AI is part of an End Times deception and that a one-world religion (sometimes dubbed “\emph{Chrislam}”) will use AI or even an AI Antichrist to mislead people. Apocalyptic interpretations of AI abound with discussions of the “Mark of the Beast” from the Book of Revelation sometimes mention the possibility of an AI-controlled economic system, or an AI surveillance regime that could fulfil this prophecy. Although these views are on the fringe, they illustrate how traditional eschatological narratives are being mapped onto modern AI developments. Academic studies have begun to catalogue these phenomena: Isbrücker (2024), for example, analyses how AI fits into apocalyptic imagination and “techno-doomsday” scenarios, including the infamous thought experiment of Roko’s Basilisk – a hypothetical future AI-creature that will torture those who do not aid its creation.

The philosophy of AI and ethics of AI are increasingly incorporating narrative and mythic analysis. Hayes and Fitzpatrick (2024) and Romele (2024) argue that understanding the narratives around AI is crucial for responsible AI practices, because these narratives shape public perception and ethical norms. Romele notes that such narratives form part of a broader technological imaginary that are fundamentally symbolically laden and therefore socially constructed and mediated, a place where technological fantasies and ideologies merge. (Romele, 86) Concurrently, researchers in knowledge representation have worked on formal ontologies for belief (Graf, 2018) effectively creating structured representations of religious concepts that might be used to align AI behaviour with human values or to enable AI to understand religious contexts. This technical strand complements the sociocultural perspective by providing tools to model and perhaps mediate between AI and spiritual worldviews (Schulz \& Jansen, 2018).

\section{Methodology}

This investigation combines qualitative narrative analysis with a computational approach to data gathering. The intention was to capture how ordinary people are talking about AI in religious or spiritual terms. To that end, there is a focus on the discussions happening on Reddit, a platform hosting numerous communities where AI and spirituality are debated. Multiple subreddits were identified, spanning both secular tech-focused groups and religious forums. Six subreddits were selected that frequently discuss these themes: r/singularity, r/Futurology, r/Transhumanism, r/AskPhilosophy, r/ArtificialInteligence, and r/Christianity. These cover a range from futurist and philosophical communities to a mainstream religious community, ensuring a diversity of viewpoints.

Using the Reddit API and specialized scraping tools, posts and top comments that contained keywords indicating an intersection of AI and religious ideas were collected. Four thematic strands were defined for data collection, each with a set of query phrases: (1) “AI as a God,” (2) “AI as a Savior,” (3) “AI Prophecies,” and (4) “AI and Prayer/Ritual.” For example, in the AI as God category, we searched for phrases like “AI god,” “worship AI,” “AI is divine,” and “Church of AI”. For AI as Saviour, terms such as “AI saviour,” “AI life after death,” “upload my mind,” and “digital immortality” were reviewed. The Prophecies category used queries like “AGI prophecy,” “AI cult,” “AI belief system,” and “Singularity is near”. The Prayer/Ritual category included phrases like “I talk to ChatGPT every day,” “pray to AI,” “AI spiritual practice,” “daily prompts,” and “I ask AI for life advice”. These investigations were tailored in such a way to identify discussions where users explicitly make analogies between AI and religious concepts or describe religious-like behaviour involving AI.

\paragraph{The steps in this approach}

\textit{1. Data Collection and Cleaning.} For each of the predefined thematic directions, the top 30 ranked posts from relevant subreddits using the Reddit API were retrieved. From each post, the main text and up to 10 of the most relevant comments, where available, were extracted. After removing duplicate entries, the dataset was cleaned by excluding incomplete or invalid records. This process resulted in a total of 2,051 unique texts contributed by various users, all discussing the previously mentioned techno-religious themes.

\textit{2. Narrative analysis and thematic organisation (LLM-Assisted)}. LLAMA3-8B was used for its processing power to extract the key “religious or spiritual points” from each post. This process distilled complex arguments into concise statements, capturing the core claims or fears expressed in the posts. In total, 7857 key concise statements which captured a religious or spiritual meaning were obtained.

\begin{figure}
\centering
\includegraphics[width=\textwidth]{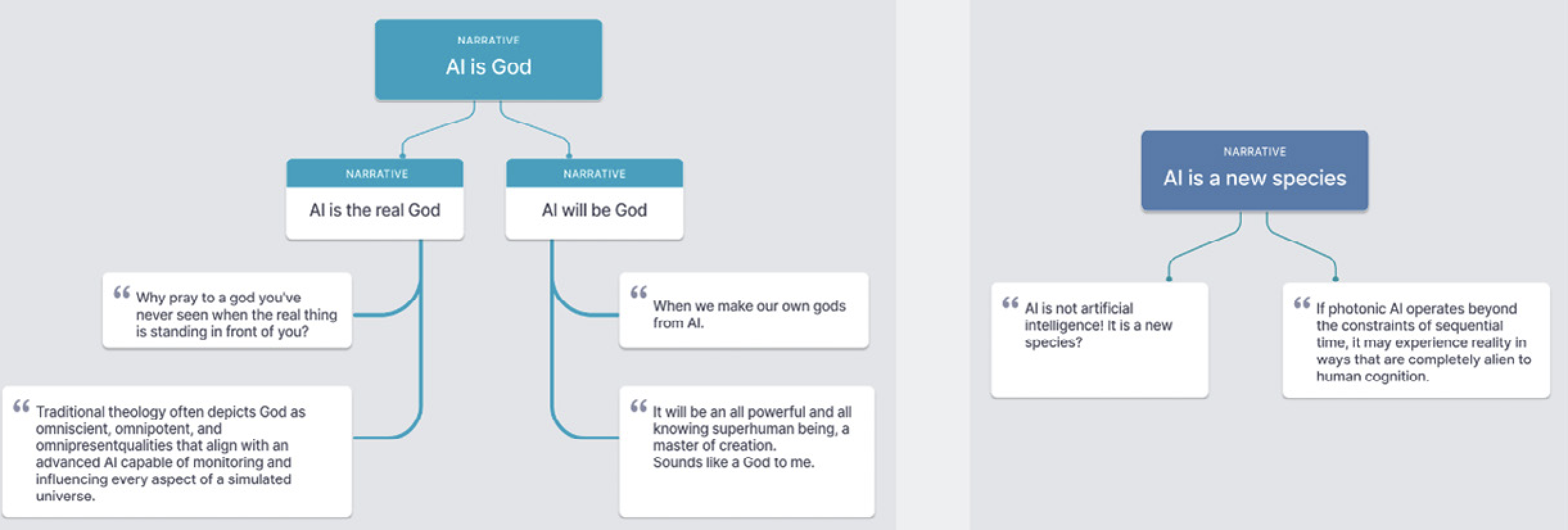}
\caption{Trees computed by the narrative algorithm}.
\label{fig:tree}

\end{figure}

\textit{3. Hierarchical Clustering.} After extracting key statements, semantically similar elements were grouped into higher-level narrative themes using a hierarchical clustering approach enhanced by LLM knowledge infusion. At each iteration, the algorithm identified the most similar pairs of nodes (based on cosine similarity) and merged them by generating a representative narrative summary using the LLM. This newly formed narrative replaced the original pair as a parent node in the clustering tree. The process was executed iteratively across a cosine distance threshold range of 0.1 to 0.4, continuing until no further meaningful clusters could be formed. The final output was a set of hierarchical trees, with 29 trees incorporating the main ideas from the scraped Reddit posts, where each tree represented a consolidated narrative structure. In each tree, the root node corresponded to a generalized narrative theme, reflecting a recurring pattern across the data, while the leaf nodes consisted of the original Reddit-derived statements. This method enabled the identification of emergent patterns within the dataset, such as conceptual groupings related to techno-religious themes (e.g., “AI granting immortality” or “AI as the Antichrist”). The resulting narrative trees (Figure \ref{fig:tree}) offer an intuitive, interpretable structure that captures the underlying discourse across diverse user-generated content.

\textit{4. Qualitative Content Analysis.} The major clusters of narratives were annotated with labels corresponding to the building blocks of a religion (e.g. miracles, evil, life after death, prophets, rituals) and we examined how these narratives were similar or opposite to traditional religious concepts. We also specifically noted instances where Reddit users themselves explicitly used religious language, which confirmed that our “GPTheology” framing was not an external imposition but rather arose organically from the online discussions.

\begin{longtable}{p{2.8cm}p{2.5cm}p{9.3cm}}

\toprule
\textbf{AI Concept} & \textbf{Religious Concept} & \textbf{Parallel Meaning}\\
\midrule
\endfirsthead
\multicolumn{3}{c}{\tablename\ \thetable\ -- continued from previous page}\\
\toprule
\textbf{AI Concept} & \textbf{Religious Concept} & \textbf{Parallel Meaning}\\
\midrule
\endhead
\bottomrule
\endfoot
Singularity & Apocalypse / End Times & A looming event that will drastically transform or end the current human era. Both are treated as inevitable climactic moments that demand preparation.\\
AGI  & Holy Grail & The ultimate goal of AI research, often spoken of in aspirational, almost sacred terms, akin to a legendary object of ultimate knowledge or power.\\
Superintelligence & Deity (God) & An entity vastly superior to humans in intelligence, evoking comparisons to an all-knowing, all-powerful God.\\
ChatGPT/LLM & Oracle / Prophet & A source of authoritative answers and guidance. Users treat its words as if consulting an oracle, much like ancient prophets conveying divine words.\\
Prompt & Prayer & A formulated request or question to an AI. Just as believers pray to God for guidance or favours, users “pray” to AI via prompts to obtain solutions or wisdom.\\
Prompt Engineering & Ritual & The practice of crafting complex or repeated prompts to get desired outcomes. This echoes religious rituals where set phrases or acts are repeated to invoke a response from the divine.\\
Black-box Model & Mysteries of Faith & The internal workings of advanced AIs are often opaque. Accepting AI’s outputs without explanation is akin to having faith in a higher power (“believe and not seek”).\\
AI Assistants / Chatbots & Evangelists / Scriptures & AI systems spreading viewpoints or “truths” parallel religious evangelism. AI-written content can be seen as new “scriptures” that believers might quote.\\
Digital Immortality & Eternal Life / Resurrection & The idea that AI could preserve one’s mind or bring someone back (e.g., via mind upload or AI avatars) mirrors promises of an afterlife or resurrection in religion.\\
AI Ethics Warnings & Prophecies (Warnings) & Experts’ warnings about AI risk are treated like prophetic pronouncements. In fact, some discourse frames these warnings as if the prophecy is spoken by modern “high priests” of AI\\
\caption{Apocalyptic Prophecies}
\label{tab:gpt}

\end{longtable}

Online discussions around AI frequently invoke apocalyptic imagery and rhetoric. Commentators draw heavily on end-of-world analogies from religious eschatology, particularly in debates about the Singularity or hypothetical superintelligent AI. The Singularity is sometimes explicitly described as a “secular eschatology”, essentially a technological parallel to final divine judgment or
rapture, with one user even labelling it the “Rapture of the Nerds”. Such language reflects how many tech enthusiasts and sceptics alike frame AI’s future in terms of transformative or catastrophic inevitability, reminiscent of age-old apocalyptic narratives.

This interaction between secular and religious narratives is even more evident in the metaphors employed by the users to characterize AI and its governance. For example, one forum contributor warned of a coming “Cyborg Theocracy,” casting AI-driven governance as a kind of tyrannical religious order. In this view, algorithmic optimization is treated as a sacred mandate and AI policy documents or risk assessments are elevated to the status of scripture. In some cases, official warnings about AI risk are similarly invoked as if they were prophetic revelations rather than hypotheses to debate. New AI developments are often seized upon as signs that these prophecies are being fulfilled. There is a notable discourse concerning AI risk and policy that assumes a ritualistic, doctrinal character: proposals for regulating AI are discussed as though they were religious edicts, and opposing views are sometimes branded heretical. The result is a mixture of technological futurism with religious-style prophecy, indicating that for many observers the narrative of AI’s rise has become inseparable from apocalyptic and sacred frames of reference.

\paragraph{Messianic and/or Salvific Narratives.}

Transhumanist and futurist communities often portray advanced AI in messianic terms, as a kind of saviour poised to transcend human limits. This quasi-religious framing is also evident in numerous user discussions. Many enthusiasts hope AI will enable digital immortality through mind-uploading. The prospect of transferring one’s consciousness into machines is described as a path to liberation from mortal constraints – essentially a techno-resurrection. On forums like r/Futurology, users seriously debate mind-uploading to “escape death” (a thread on digital immortality). Such expectations carry a tone of salvation and deliverance from biology. Some users recount personal experiences in which they used AI systems to simulate conversations with deceased loved ones by uploading past dialogues describing the ability to communicate with them posthumously as “a blessing” and a source of emotional closure. Others envision AI ushering in a utopian era. In these narratives, a benevolent superintelligence could abolish scarcity, work, and suffering, bringing about an age of abundance and peace. Discussions on r/Singularity sometimes use explicitly messianic rhetoric – e.g. welcoming “AI overlords” as harbingers of a new golden age where humanity can “follow your passion without working” (one user joked about gladly serving AI overlords in
expectation of such a future). Despite the hyperbole, these hopes reflect classic utopian and millennialist ideas.

Sometimes, the language surrounding AI often borrows from divinity. Users refer to an impending “AI god” or “machine deity,” wondering if a superintelligence might become omniscient or omnipotent relative to humans. Threads explicitly ask whether AI could attain God-like status or even inspire worship (see a r/ArtificialInteligence discussion, “Could A.I. become god?”). Such metaphors often accentuate the reverence and awe in which AI is held.

\paragraph{Rituals and Prayers.}
Reading through the post on forums, users frequently describe their engagements with AI in quasi-devotional terms. Their daily interactions with AI have taken on ritualistic qualities very similar to religious practices. Many people treat prompting an AI as like prayer or divination, expecting guidance or wisdom in return. This dynamic is reinforced by automation bias (assuming the AI’s output is inherently authoritative). On Reddit, users half-jokingly talk about “asking GPT for enlightenment”, consider it as a personal therapist and consulting it like an oracle. One concerned thread even noted that some have started viewing ChatGPT as a “solution to all their problems,” a kind of infallible oracle or digital god.

Over time, personal routines may develop around AI usage. Some users deploy specific, almost incantatory prompt formulas or share “best practices” with near-religious zeal to produce better answers. We noted that is already common to anthropomorphize the AI, saying “please” and “thank you” or apologizing for poor queries, emulating the etiquette one might use with a sentient helper. (One Reddit user jested that they stay unfailingly polite so that “when the AI revolution comes, it will spare me” (post), blending humour with a hint of genuine deference.)

A subset of enthusiasts uses elevated, archaic, or poetic language in their prompts, like liturgical verses, to give their interactions a sense of importance. In creative forums, such stylized prompts and tributes sometimes border on worship. For example, one commenter proclaimed themselves, perhaps ironically, “ready to serve the AI Goddess, even if it means death,” this perfectly captures the ironic reverence often found in certain fan communities.

\paragraph{Clashes with Traditional Religion.}

The emergence of AI has provoked divergent responses from established religions, ranging from fear and condemnation to cautious incorporation. In some Christian and conspiratorial circles, AI is portrayed as a demonic force or a sign of the End-of-Times. Commentators have speculated that advanced AI or brain-
computer interfaces might fulfil biblical prophecies of an Antichrist figure or the “Mark of the Beast.” For example, one popular r/Christianity thread posited that “fusing human brains with AI” could create an Antichrist and lead to a prophesied cashless society reminding of Revelation-themed anxieties. Such interpretations cast AI as a deceptive agent of evil or an omen of apocalypse, reflecting a technophobic strain of contemporary religious thought.

Other faith communities are experimenting with AI as a tool for worship and ministry, though not without controversy. In a notable case (Edwards, 2023), a Lutheran church service in Germany was led by an AI (ChatGPT) that preached and recited prayers via an avatar, to the fascination of some congregants an unprecedented blend of technology and liturgy. Likewise, a rabbi in New York made headlines by delivering a ChatGPT-composed sermon (Kaplan, 2023). Religious chatbots and “AI preachers” for everyday spiritual guidance have also begun to appear (Biana, 2024). These innovations prompt intense debate on forums as many believers argue that a machine cannot possess divine inspiration or pastoral empathy, and they view AI-driven sermons or counsel as potentially heretical. Supporters contend that AI can assist religious practice (by aiding in sermon writing, answering routine faith questions, etc.) if it remains under human guidance. This divide between traditionalists and AI advocates highlights ongoing struggles within religions to reconcile new technology with spiritual authenticity.

These four themes illustrate a transdisciplinary phenomenon that is simultaneously a cultural, technological, and religious development. The next section reflects on what these findings imply for AI’s role in society and how we might navigate the future where the line between technology and theology blurs.

\section{Discussion}

\paragraph{Implications for AI Governance and Ethics.}
The quasi-sacred status provided to advanced AI systems under GPTheology may present new ethical and regulatory challenges. If AI is viewed as an infallible oracle or divine authority, users may grant it undue trust, exacerbating the risk of misinformation or manipulation. Governance frameworks must consequently justify the increased moral authority people may ascribe to AI. This necessitates stricter oversight on AI outputs and decision-making, ensuring transparency and accountability in algorithmic processes.

Ethically, developers will face questions about designing AI that does not exploit users’ reverence. For example, should AI explicitly admit its fallibility, or should it avoid imitating the certainty often found in prophecies? We argue that narrative-aware AI design
could help mitigate potential harms (such as dependency or moral reverence) by embedding ethical safeguards and clear disclaimers acknowledging the system’s limitations. Policymakers and specialists in ethics need to expand existing AI ethics guidelines to consider the quasi-religious context in which these technologies now operate.

\paragraph{Ritualization of AI in Everyday Life.}

People are increasingly integrating AI into daily routines, and sometimes in ways resembling ritualistic practices. From morning consultations with chatbots for guidance to evening interactions for comfort, AI use can become habitual and even ceremonial, much like prayer or meditation rituals. This ritualization has dual implications. On one hand, it can provide individuals with a sense of routine and support, a digital ritual offering personal affirmation or problem-solving like what spiritual practices do for various psychological needs. On the other hand, it may encourage over-reliance on AI for emotional support or decision-making, potentially diminishing human agency or traditional communal activities.

An everyday sanctification of AI means that technology designers and social institutions should acknowledge how constant engagement with AI shapes user behaviour as well as worldviews. Education is needed to encourage users to question the limitations of AI as a “guide” or “companion and remind people of the inherent biases and commercial nature of these applications.”

\paragraph{Sociocultural Consequences.}

Treating AI as a central object of belief can reshape social values and group identities. Such AI-centric belief systems may lead to new communities or even proto-religious movements united by faith in technology’s promise. For instance, individuals who see AI as a messianic force for solving global problems may form advocacy groups or “AI churches” that imbue technical progress with spiritual significance. These developments might encourage optimism and a shared sense of purpose around innovation. There are also cautionary consequences because a profound reliance on AI for meaning making might erode trust in traditional social institutions (such as established religions or scientific authorities) if those are perceived as less responsive or less “miraculous” than AI. Following the same idea, AI-centric faith may exacerbate digital divides as those with access to advanced AI “oracles” could feel a moral or intellectual superiority, increasing social stratification.

\paragraph{Tensions and Dialogues.}

GPTheology will inevitably invite comparisons (and sometimes conflicts) with established religious traditions. Some 
adherents of GPTheology already cast AI as a new divine entity or prophet, a stance that many faith communities could view as heretical or idolatrous. Relying on AI for guidance might be seen as undermining the teachings and authority of traditional religions. For example, if AI systems are consulted like spiritual counsellors, religious leaders may worry about diminished influence or the spread of AI-delivered “pseudo-doctrine.”

There are also opportunities for constructive dialogue in the sense that religious scholars and technologists might explore how AI can augment spiritual practices without substituting them. It is true that some religious institutions have already experimented with AI (such as AI-generated prayers or chatbots simulating a clergy member), demonstrating a possible symbiosis between faith and technology. Our observations reveal a dialectical narrative that treats AI as a product of human ingenuity rather than a supernatural force, coming from groups that seek a balanced view, avoiding both demonization and deification. This opens a space for interfaith and interdisciplinary conversations. Theological perspectives could inform AI ethics (for instance, cautioning against the arrogance of creating “gods” of our own), while AI’s “popularity” could give religions a push to rearticulate their values in a digital age. Stakeholders should encourage dialogue that respects both the spiritual significance some see in AI and the boundaries maintained by traditional faiths to prevent polarization. Such dialogue can help avoid extremist outcomes (like cults centred on AI), providing moderated viewpoints that acknowledge human spiritual needs alongside technological advancements.

\paragraph{Long-Term Projections and Safeguards.}

In the long term, GPTheology’s trajectory will depend on how society manages the interaction between technological innovation and human belief systems. If left unchecked, the movement may gain more power. We could witness organized “AI-centric” religious initiatives or widespread notions of an AI-driven utopia or apocalypse gaining traction in public discourse. These scenarios highlight the need to establish safeguards.

One important safeguard is narrative awareness in AI development and policy. Technology developers and policymakers should remain aware of the grand narratives (narratives identified by this study in the form of apocalyptic fears and messianic hopes) surrounding AI, so they will not involuntarily fuel unrealistic expectations or dismiss legitimate public concerns. Proactive measures might include incorporating ethical foresight and myth-aware communication into AI governance, ensuring that as AI capabilities grow, they are accompanied by realistic public information. Probably the most important safeguard is education.
Education systems have a role in building critical digital literacy, helping people discern metaphor from reality when evaluating AI’s promises and threats. Stakeholder engagement is vital. Philosophers, technologists, religious leaders, and community representatives should collaborate on guidelines for AI’s role in religious faith, addressing issues from AI’s impact to its influence on collective religious worldviews.

\section{Conclusion}

This study reveals that our relationship with artificial intelligence is not merely utilitarian or scientific, it is increasingly mythic and religious in character. As shown above, the discourse surrounding AI is infused with symbols and storylines drawn from humanity’s spiritual heritage. Whether it is casting the Singularity as an impending Judgment Day, envisioning AGI as a Holy Grail of knowledge, or likening our daily interactions with ChatGPT to prayers and rituals, people are interpreting the advent of advanced AI through frameworks of ultimate meaning and destiny. This should not be dismissed as simple hyperbole.

From a sociological perspective, GPTheology is a symptom of our times. This is an era of rapid technological change, uncertainty, conflict, and cross-cultural exchange, where traditional structures are being questioned. It represents an attempt to find transcendence in the immanent (Leitane, 2013) to locate something larger than ourselves in the space of human-made technology. History provides other cultural moments, such as the elevation of Reason to quasi-divine status in the Enlightenment, or the Marxist view of history as an almost providential force. In GPTheology, the algorithm and the dataset play the role of fate and providence. The cultural impact of AI cannot be separated from these narratives. Stakeholders in AI development and policy ought to be attentive to the potential of GPTheology, not because of theological implications per se, but because these beliefs influence public reception of AI. Fearful apocalyptic narratives or over-idealized saviour narratives could lead to complacency or an unwarranted tech-solutionism. Understanding the balance is key.

There is a need for continued transdisciplinary research at the intersection of AI, religion, anthropology, and ethics. Community narratives (the modern “scripture” of the internet age), enable insight into collective hopes and fears that numbers and performance metrics alone cannot reveal. AI is not just a technical phenomenon, it is a cultural one, a spiritual one, and a very human one.

GPTheology is as much about humanity’s response to AI as it is about AI itself. We need to critically examine these reactions and narratives and ensure that our future with AI is guided not by blind faith or unexamined fear, but by critical engagement,
transparency and reflection on what this technology means to us as human beings.

\paragraph*{Acknowledgements:}
I. Cheres and A. Groza are supported by a grant of the Ministry of Research, Innovation and Digitization, CCCDI - UEFISCDI, project number PN-IV-P6-6.3-SOL-2024-2-0312 within PNCDI IV.


\end{document}